# Double-Hashing Algorithm for Frequency Estimation in Data Streams

Nikita Seleznev
Capital One
Cambridge, Massachusetts
nikita.seleznev@capitalone.com

Senthil Kumar
Capital One
New York, New York
senthil.kumar@capitalone.com

C. Bayan Bruss
Capital One
McLean, Virginia
bayan.bruss@capitalone.com

## ABSTRACT

Frequency estimation of elements is an important task for summarizing data streams and machine learning applications. The problem is often addressed by using streaming algorithms with sublinear space data structures. These algorithms allow processing of large data while using limited data storage. Commonly used streaming algorithms, such as count-min sketch, have many advantages, but do not take into account properties of a data stream for performance optimization.

In the present paper we introduce a novel double-hashing algorithm that provides flexibility to optimize streaming algorithms depending on the properties of a given stream. In the double-hashing approach, first a standard streaming algorithm, such as count-min sketch [1], is employed to obtain an estimate of the element frequencies. This estimate is derived using a fraction of the data stream and allows identification of the heavy hitters. Next, it uses a modified hash table where the heavy hitters are mapped into individual buckets and the rest of the stream elements are mapped into the remaining buckets. Finally, the element frequencies are estimated based on the constructed hash table over the entire data stream following any streaming algorithm, such as count-min sketch.

We demonstrate on both synthetic data and an internet query log dataset that our approach is capable of improving frequency estimation due to removing highly frequent elements from the hashing process and, thus, reducing collisions in the hash table. Our approach avoids employing additional machine learning models to identify heavy hitters and, thus, reduces algorithm complexity and streamlines implementation. Moreover, because it is not dependent on specific features of the stream elements for identifying heavy hitters, it is applicable to a large variety of streams.

In addition, we propose a procedure on how to dynamically adjust the proposed double-hashing algorithm when frequencies of the elements in a stream are changing over time.

## 1 INTRODUCTION

Estimation of the frequency of items is one of the fundamental tasks in data stream summarization. Requirements to analyze large data streams led to a development of data sketching techniques. The requirement to analyze data streams arises in many scenarios such as monitoring of the internet traffic with a router or a search engine monitoring of internet queries.

Data sketching algorithms are sublinear space data structures for summarizing data streams [2]. A sketch $C(X)$ of data $X$ with respect to some function $f$ is a compression of $X$ that allows one to approximately compute $f(X)$ while having only $C(X)$. In application to streaming algorithms it is important to update a sketch $C(X)$ in real time as the data stream items come and $X$ is updated. Sketching algorithms require much less memory compared to the data stream and often provide answers in constant time. The provided answers are approximate, but allow for theoretical guarantees on error bounds.

An important class of data sketching algorithms is based on hash functions, such as the count-min sketch [1], a classic algorithm used for frequency estimation. Count-sketch is another hashing-based algorithm used for frequency estimation [3]. Besides frequency estimation, sketching algorithms have found applicability in other areas. For example, the Bloom filter, a space-efficient sketch, is commonly used to test whether an element is a member of a set [4].

A class of learning-augmented algorithms has recently been proposed [5]. Within a learning-augmented approach an algorithm learns relevant parameters from the input data and uses these parameters to improve its performance. This general framework has been applied to various problems, including identification of most frequent items in the stream, which is often referred to as the heavy hitters problem.

## 2 RELATED WORK

Classical sketching algorithms allow formal guarantees on their performance as a function of the algorithm parameters, but they do not utilize properties of the data streams to optimize algorithm performance.

A recently introduced learned sketch concept allows optimization of classical sketching algorithms for frequency estimation in data streams [5]. Specifically, the learned sketch approach attempts to reduce frequency estimation errors in classical hashing-based sketching algorithms by customizing their hash tables using data stream patterns. The most frequent items are assigned to their own unique buckets to minimize collisions in the shared hash table buckets. Identification of heavy hitters in the stream is an important part of a learned sketch algorithm Hsu et al. proposed and used machine-learning models for this purpose [5]. In the present paper we compare our results to their method among other baselines.

The machine learning-augmented frequency estimation algorithm was further improved in terms of frequency estimation errors by optimizing the training strategies and loss function for the machine-learning part of the algorithm [6].

The learning-augmented algorithms have also found applications in other areas, such as memory caching. The developed framework augmented online algorithms with a machine-learned oracle [7]. This learning-augmented approach was further improved by providing near-optimal bounds for online caching [8]. Yet another implementation of the learning-augmented approach for online caching was developed to optimize its performance in the presence of structured inputs [9].

## 3 DOUBLE HASHING FREQUENCY ESTIMATION ALGORITHM

First, we would like to review standard data sketching approaches, such as count-min sketch. The count-min sketch is a sublinear data structure for summarizing data streams. Among other applications it allows a quick estimation of the item's frequency in the stream.

An underlying principle of a quick estimate of the item frequencies with a sublinear data structure involves utilization of hash functions.

A hash function $h$ allows mapping of a set $\{1..n\}$ to a set $\{1..w\}$. The set $\{1..w\}$ is referred to as the hash table. In application to a large data stream processing it is common that $n > w$. In this case some items of the input set will be mapped in the same elements of the hash table, which is referred to as collision. A good hash function should be very fast to compute and it should minimize collisions. A review of practical hash functions is given in [10].

For estimation of items frequency the following configurations can be used:

**A single hash function:**

A single hash function maps each item of the input stream into the hash table: $h: \{1..n\} \to \{1..w\}$. The algorithm requires maintaining a single array $C$ of the size $w$ to keep track of the item's frequencies. All elements of the input array are initially set to 0. When a next stream element arrives, the hash function, $h$, maps the arrived element $i$ into the hash table element $m = h(i)$ and increments its value $C[m]$.

At the end of the stream the value of the hash table for an element $m$ is given by the following equation:

$$C[m] = \sum_{j:h(i)=m} f_j$$

where $f_j$ is the frequency of the element $j$ in the data stream.

Then the estimate $\hat{f}_i$ of the $i-th$ element frequency can be obtained from the hash table as: $\hat{f}_i = C[h(i)]$. It is evident that this conditions is always satisfied: $\hat{f}_i \geq f_i$

The error in the frequency estimate is inversely proportional to the size of the hash table as the number of collisions grows with decreasing table size.

**Multiple hash functions:**

Several approaches that allow further error reduction in the frequency estimation are based on utilization of multiple hash functions. Specifically, the count-min sketch uses $d \geq 2$ distinct hash functions. Correspondingly, this approach requires a hash table $C$ of size $w \times d$. When $i-th$ element of the stream arrives it is mapped by the $k-th$ hash function to the element $C[h_k(i), k]$ and its value is incremented.

Double-Hashing Algorithm for Frequency Estimation

At the end of the stream the frequency estimate $\hat{f}_{i,k}$ of the $i-th$ element is produced for each hash function $k$ as in the single hash function approach. Then, in the count-min sketch approach the smallest frequency estimate is chosen as the best estimate: $\hat{f}_{i,cm} = min(\hat{f}_{i,k})$. For the count-min sketch condition $\hat{f}_{i,cm} \geq f_i$ is also always satisfied.

The error in the frequency estimate is decreased as the number of hash functions increases and/or the number of bins available for each function increases. Theoretical guarantees can be provided on the algorithm error bounds [1].

Other sketching algorithms utilizing multiple hash functions, such as count-sketch [3], can also be used within the double hashing framework proposed below.

**Learned Sketch**

Sketching algorithms presented above do not take into account properties of the analyzed data stream. These algorithms treat all stream elements equally regardless of their frequency.

Errors in the frequency estimation obtained with sketching algorithms arise due to collisions in the hash tables. Intuitively, items that are dominant in the stream cause more significant error in the frequency estimation when they are involved in collisions. Thus, if a subset of elements has substantially higher frequency in the stream compared to the rest of the elements it may be beneficial to allocate unique buckets for these frequency elements even at the expense of some reduction of the remaining hash table size.

The most frequent items in the stream are referred to as the "heavy hitters". The $k-heavy\ hitters$ are defined as the stream elements with indices $i$ whose frequencies $f_i$ satisfy the following criterion: $f_i > \left(\frac{1}{k}\right)\sum_{i=1}^{n} f_i$. The number of elements which can be strictly larger than a γ-fraction of this sum must be strictly less than $1/\gamma$, and, thus, there can be at most $k-1$ heavy hitters in the stream.

The approach that optimizes the sketch performance based on the properties of a particular stream by assigning unique buckets to the heavy hitters is referred to as the learned sketch. Utilization of machine-learning models was proposed as a means to identify heavy hitters within the data stream [5], [6].

**Double Hashing Algorithm**

In the present work we are proposing a new approach called double hashing for frequency estimation in data streams. We build upon the learned sketch framework that enhances performance of the commonly used sketching algorithms by identifying the heavy hitters and tailoring the algorithm to the stream properties.

In the proposed approach the identification of the heavy hitters is carried out by applying a sketching algorithm, such as count-min sketch. Initial identification of heavy hitters is done on a fragment of the stream and is referred to below as the first algorithm pass. Mere identification of the heavy hitters does not require algorithm optimization and can be done with a standard sketching approach using limited space.

Once a list of heavy hitters is identified we employ the learned sketch framework where the heavy hitters are assigned their own unique buckets to prevent collisions due to frequent items in the rest of the hash table. The rest of the stream elements are mapped into the remaining buckets, which number is defined by the available memory space. Finally, the element frequencies are estimated based on the constructed hash table over the entire data stream following any streaming algorithm, such as count-min sketch.

The algorithm tuning step optimizes the heavy hitters cutoff and the number of hash functions based on a fragment of the stream. Then, the optimized algorithm can be used to produce an optimum estimation of the element frequencies in the stream.

The double hashing algorithm avoids employing additional machine learning models to identify heavy hitters and, thus, reduces algorithm complexity and streamlines implementation. Implementation of a sketching algorithm can be reused for the heavy hitter identification.

**Continuous Optimization**

Because sketching algorithms have low memory requirements and often provide answers in constant time the double hashing algorithm can be used throughout the stream analysis for continuous algorithm optimization.

At any point in time the double hashing algorithm carries information about the current frequencies of the stream elements. The frequency estimates can be produced based on the counts stored in the hash table. If new heavy hitters emerge they can be assigned their own unique buckets. If some of the previously identified heavy hitters become less frequent in the stream their frequency counts can be hashed into the rest of the hash table and their unique buckets can be freed up for new heavy hitters. Such monitoring allows continuous optimization of the algorithm to the data stream without making any *a priori* assumptions on its properties.

A pseudo-code for the double hashing algorithm is given below as Algorithm 1. A block-diagram for our algorithm is shown in Figure 1.

$M$ is the total space used by the algorithm, $B_c$ is the number of buckets used by the sketch, $B_u$ is the number of unique buckets assigned to the heavy hitters, $d$ is the number of hash functions used by the sketch $C$, $d_{hh}$ is the number of hash functions used by the sketch $C_{hh}$, $(i,f)$ is a key-value pair describing element $i$ with frequency $f$ in the stream, $S_t$ is the fraction of the stream used to determine heavy hitters, $S$ is the data stream.

---

**Algorithm 1** Double-Hashing Algorithm

---

**input**: $M$, $d$, $S$, $S_t$

Initialize sketch $C$ with $B_c$ bucket, $d$ hash functions

**for** each pair $(i,f)$ in $S_t$:

  provide $i$ as an input to $C$

**end for**

Calculate frequency estimates $\hat{f}_i$ from $C$

Determine heavy-hitters list $HH$ from $\hat{f}_i$

Initialize sketch $C_{hh}$ with $B_c - B_u$ buckets, $d_{hh}$ hash func.

**for** each pair $(i,f)$ in $S$:

  if $i$ is in $HH$:

    increment $B_u(i)$ by $f$

  **else**

    provide $i$ as an input to $C_{hh}$

  **end if**

**end for**

Determine frequencies of $HH$ from $B_u$

Determine frequencies of non $HH$ $\hat{f}_i$ from $C_{hh}$

---

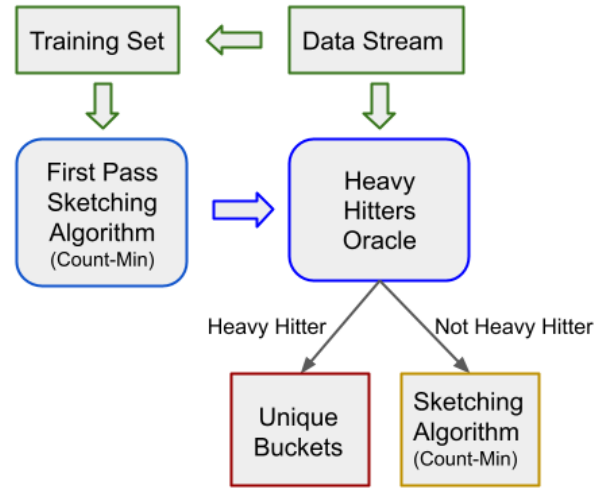

Figure 1: **Block-diagram for the double-hashing algorithm.**

## 4 EXPERIMENTS ON SYNTHETIC DATA

Zipf's law is an empirical law that was found to satisfactorily describe many types of data in the physical and social sciences [11]. Zipf's law postulates that the rank-frequency distribution has an inverse relationship. Let $N$ be the number of elements in a dataset, $k$ be the rank of the elements, and $s$ be the exponent value characterizing the distribution. Then the Zipf's law can be formulated as:

$$f(k;\ s,\ N) = \frac{1/k^s}{\sum_{n=1}^{N}(1/n^s)}$$

where $f(k;\ s,\ N)$ is the normalized frequency of the element of rank $k$.

Initially Zipf's law found applications in linguistics where it was noticed that the frequency of words is inversely proportional to their rank in a corpus. The word frequencies in human languages can be reasonably well modeled by a Zipfian distribution with $s = 1$ [12]. Zipfian distribution was also applied to represent item frequency distribution in data streams [5], [6]. Following these studies we adopt Zipfian distribution to generate streaming data for our synthetic experiments. Figure 2 shows probability mass function (PMF) of Zipfian distribution for different values of the exponent $s$.

We generated synthetic data with Zipfian distribution with $s = 0.7$ for 140,000 unique stream elements. This setting approximates the number of unique internet queries during one day in the AOL dataset that we use later for testing.

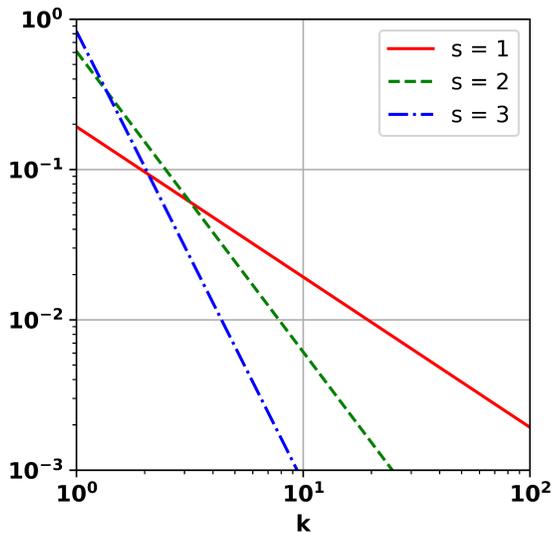

Figure 2: **Probability mass function of Zipfian distribution for different values of the exponent, $s$ as a function of rank, $k$, for $N = 100$.**

We compare performance of the proposed double-hashing algorithm to the standard count-min sketch approach and to the learned sketch approach with exactly known heavy hitters. The latter baseline we refer to as the "ideal" case as it predicts the best achievable performance for a given data stream within the learned sketch approach.

The results of the test are presented in Figure 3. The average frequency error produced by different baselines is plotted versus memory space used by the algorithms. As expected, the "ideal" learned sketch provides the best performance among the baselines. The double-hashing algorithm outperforms the standard count-min approach for all memory spaces. The minimum considered memory space of 0.1 MB is allocated for the first pass of the double-hashing algorithm.

We also conducted experiments on synthetic data generated with Zipfian distribution with exponent equal to 1. The PMF of this distribution is shown as a red line in Figure 2. Higher Zipf's exponent leads to fewer and more prevalent heavy hitters in the data stream. For comparison, the items with rank equal to 1 represent close to 20% of the stream for $s = 1$ and close to 60% of the stream for $s = 2$ for $N = 100$. Thus, we expect stronger influence of heavy hitters on the algorithm performance for higher values of the exponent.

The baseline comparison on synthetic data with $s = 1.0$ is shown in Figure 4. The frequency estimation errors are lower for all algorithms as there are fewer frequent elements in the stream. Also, the double-hashing algorithm provides near-optimum performance. It is intuitively expected that it would perform better in this setting as there are fewer heavy hitters and their fraction in the stream is higher.

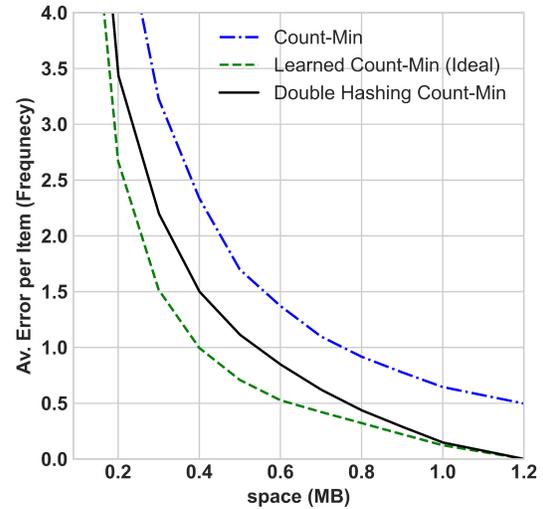

Figure 3: **Dependence of the frequency prediction error on the memory space for the standard count-min, an ideal learning-augmented account-min and double hashing algorithm. Test results generated on the synthetic data generated with Zipf's exponent = 0.7.**

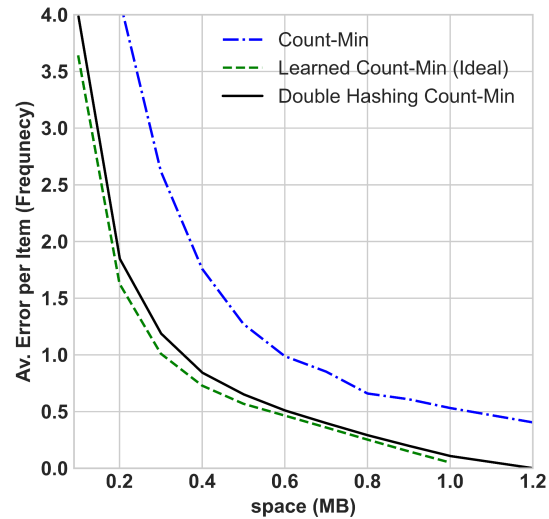

Figure 4: **Dependence of the frequency prediction error on the memory space for the standard count-min, an ideal learning-augmented account-min and double hashing algorithm. Test results generated on the synthetic data generated with Zipf's exponent = 1.0.**

Thus, heavy-hitters are easier to identify and there is a higher improvement in accuracy when heavy-hitters are isolated in their own buckets.

To sum up, experiments on synthetic data confirm that the double-hashing algorithm provides an improvement on the standard count-min approach for streams with various fractions and prominence of heavy hitters.

## 5  EXPERIMENTS ON REAL DATA

In this section we are applying the double-hashing algorithm to a real data stream collected from the internet and comparing its performance to other baselines.

**Baselines**. We compare our double-hashing algorithm with a standard count-min sketch approach, a learning-augmented algorithm described in [5], and an algorithm using an ideal heavy-hitter predictor.

The algorithm with an ideal heavy-hitter predictor uses the exact frequencies of the queries in the test set to select optimum splits between the heavy-hitters and the rest of the items. This baseline represents an ideal performance of a sketching algorithm augmented with unique buckets assignment for heavy hitters.

**Dataset:** We employ AOL query log dataset to compare our approach to the baselines. The AOL data set consists of ~21 million search queries collected from ~ 650 thousand users over a period of 90 days. The users are anonymized in the datasets. Each query is a search sentence containing one or multiple words. The dataset has approximately 3.8 million unique queries. The goal for all algorithms in comparison is to estimate the number of times a search query appears during a given period of time.

In applying our method to the dataset we follow the same procedure as in [5] to facilitate comparison with prior work. The data during the first five days is used as a training set. Our double hashing algorithm completes the first pass over the training data using a standard count-min sketch and uses the query counts to determine the heavy hitters. The accuracy requirements for this pass are less stringent as the algorithm only needs an identification of heavy hitters. Thus, it can be accomplished with limited storage.

The learning-augmented algorithm uses the first five days to train a neural network to predict heaviness of a query. The details of the neural network predictor are discussed in [5].

The sixth day is used as a validation set, and the subsequent days are used for testing. We use the 50th and the 80th days to test performance of our approach and to compare with the rest of the baselines.

The comparison is made between the standard count-min sketch algorithm, the ideal learned count-min sketch, the learned count-min sketch with a neural network, and the proposed double hashing algorithm. The figure shows dependency of an average frequency error per item on the memory space used by the algorithms.

Following [5] an assumption is made that each bucket requires 4 bytes of memory. For the case of the learned sketch and double hashing algorithms each unique bucket requires 8 bytes of memory.

Initially, the space of 0.2 MB was chosen for the first pass of the double hashing algorithm that generates a list of heavy hitters. Four hash functions were used for the first pass. The main pass of the double hashing algorithm was tested on the same memory spaces and numbers of hash function as the other baselines. The test results for the 50th day are shown in Figure 5.

It is evident that both the neural-network based learned sketch and the double hashing algorithm provide an improvement on a standard count-min sketch.

The neural network-based algorithm shows better performance for larger memory space values, while the double-hashing algorithm performs better at lower memory spaces. Specifically, it performs better for 0.2 MB of space. The ideal learned sketch outperformed all other approaches indicating that other algorithms can still be improved.

First, we increase the available memory space for the first pass of the double-hashing algorithm from 0.2 MB to 1 MB. The results for the 50th day are presented in Figure 6. The increasing space available for the first pass analysis should lead to more accurate identification of heavy hitters.

However, there is only a little improvement in the average frequency error for the double-hashing algorithm despite an increase of the available memory for the first pass. It suggests that mere identification of the heavy hitters with the count-min sketch approach can be done efficiently even with limited space.

The remaining gap between the ideal learned sketch and the other algorithms is likely due to properties of the analyzed query dataset changing from the first five days used for training and the 50th day used for testing.

Double-Hashing Algorithm for Frequency Estimation

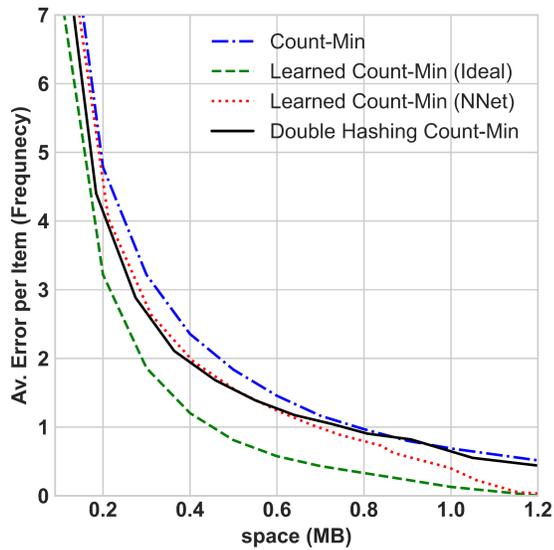

Figure 5: **Dependence of the frequency prediction error on the required space for double-hashing algorithm, standard count-min sketch, learning-augmented algorithm and an ideal case. Test results generated on the 50th day. The double hashing algorithm trained on days 0-4, and validated on day 5.**

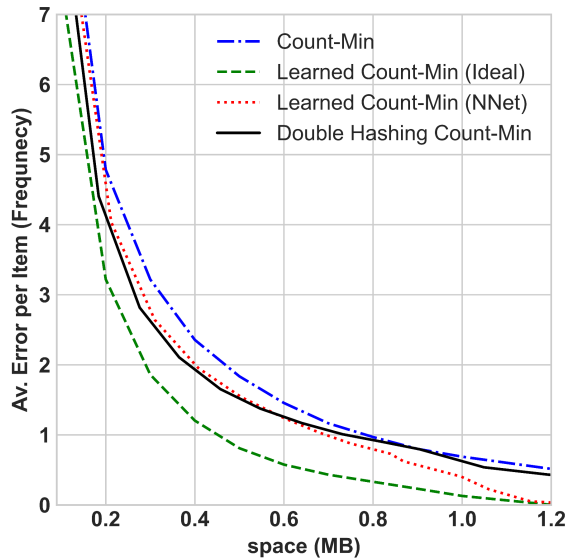

Figure 6: **Dependence of the frequency prediction error on the required space for double-hashing algorithm (increased first pass memory), standard count-min sketch, learning-augmented algorithm and an ideal case. Test results generated on the 50th day.**

We also repeat our experiments for the 80th day of testing. The results are presented in Figures 7 and 8 for the 0.2 MB and 1 MB space allocated for the first double hashing pass correspondingly. The observed patterns and conclusions are similar to the results obtained on the 50th day of testing.

One of the advantages of the double-hashing algorithm is that we do not need to train a separate model to predict heavy hitters in the stream. In our double hashing approach the prediction of the heavy hitters can be done continuously as the data stream evolves allowing to capture heavy hitters not present in the initial training set. After the initial training batch is processed the role of the first pass estimate is fulfilled by the learned sketch itself as it continuously provides an estimate of the most frequent items.

For example, in application to the AOL queries dataset the double hashing approach allows optimization of the learned sketch parameters after each streaming day. The double hashing frequency estimate for the last day of streaming is used as a validation set, while the frequency estimates for several days prior to the validation set are used as the training set.

This approach allows dynamic modification of the learned sketch parameters to reflect emerging heavy hitters, dispose of items that are not frequent anymore, and selection of the optimum bucket size and number of hashing functions reflecting the current properties of the stream.

Moreover, because this approach does not make assumptions on specific features of the stream elements for identifying heavy hitters, it is applicable to a large variety of streams.

The results of applying the double-hashing algorithm using the above scheme are shown in Figures 9 and 10. The space of 0.2 MB was chosen for the first pass of the double hashing algorithm that generates a list of heavy hitters.

Figure 9 shows the results for the double-hashing algorithm trained on days 44-48, and using validation data from day 49. From comparison with Figure 5 it is evident that the average error for the double hashing algorithm is reduced and the performance of the double-hashing algorithm compares favorably to the neural-network based learned sketch for most of the tested space range.

Application of the double hashing algorithm trained on days 74-78, and using validation data from day 79 to test data from day 80 also shows error reduction. The results are displayed in Figure 10. Comparison with Figure 7 confirms an improvement in the double hashing algorithm results. Similarly to day 50 performance of the double-hashing algorithm compares favorably to the neural-network based learned sketch for most of the tested space range except the highest values.

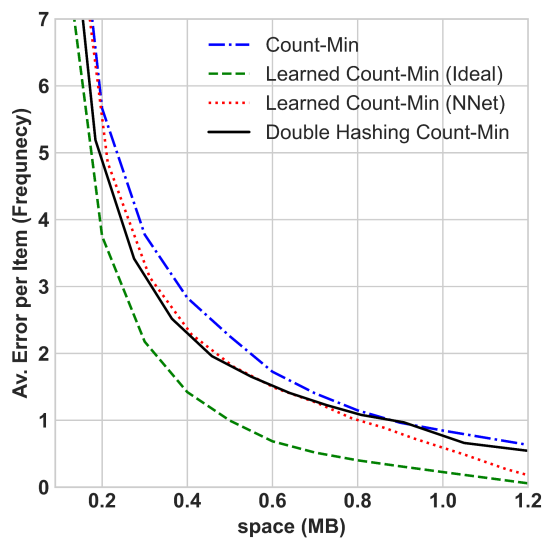

Figure 7: **Dependence of the frequency prediction error on the required space for double-hashing algorithm, standard count-min sketch, learning-augmented algorithm and an ideal case. Test results generated on the 80th day.**

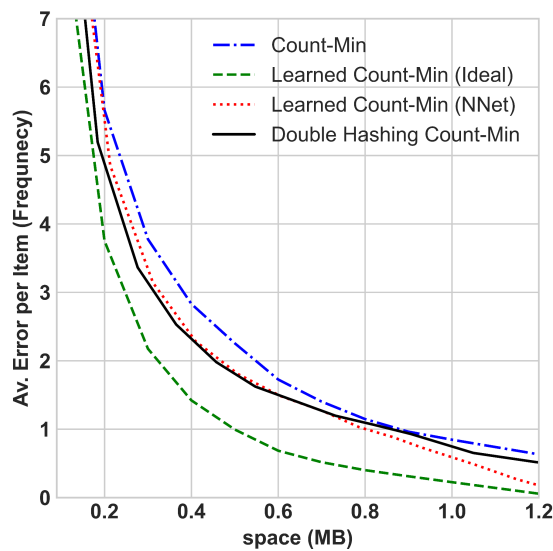

Figure 8: **Dependence of the frequency prediction error on the required space for double-hashing algorithm (increased first pass memory), standard count-min sketch, learning-augmented algorithm and an ideal case. Test results generated on the 80th day.**

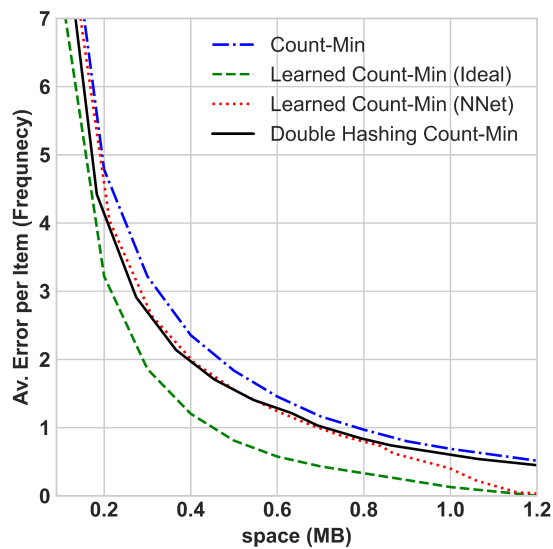

Figure 9: **Dependence of the frequency error on the required space for double-hashing algorithm, standard count-min sketch, learning-augmented algorithm and an ideal case for the 50th day. The double hashing alg. trained on days 44-48, and validated on day 49.**

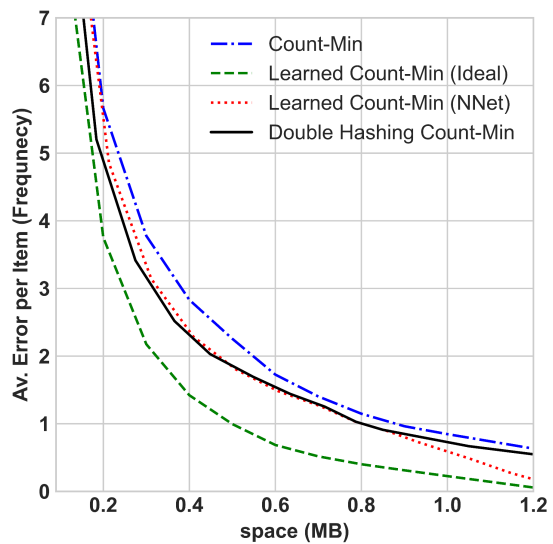

Figure 10: **Dependence of the frequency error on the space for double-hashing algorithm, standard count-min sketch, learning-augmented algorithm and an ideal case for the 80th day. The double hashing alg. trained on days 74-78, and validated on day 79.**

Numerical values of the average frequency errors for days 50 and 80 are given in Tables 1 and 2 for different space values. The double-hashing algorithm reduces frequency errors compared to the standard count-min sketch for all spaces. Training and optimizing the double-hashing algorithm on the data immediately preceding the prediction date helps to improve its performance for higher space values.

| Space | CM | DH 1 | DH 2 | CM-Ideal |
|---|---|---|---|---|
| 0.2 | 4.8 | 4.12 | 4.14 | 3.2 |
| 0.6 | 1.45 | 1.26 | 1.26 | 0.56 |
| 0.8 | 0.96 | 0.91 | 0.82 | 0.32 |
| 1.0 | 0.68 | 0.64 | 0.59 | 0.12 |

Table 1: **Average frequency error for standard count-min (CM), double-hashing with days 0-4 training and day 5 validation (DH 1), double-hashing with days 44-48 training and day 49 validation (DH 2), and the ideal learned count-min (CM-Ideal) for the 50th day prediction.**

| Space | CM | DH 1 | DH 2 | CM-Ideal |
|---|---|---|---|---|
| 0.2 | 5.63 | 4.88 | 4.84 | 3.73 |
| 0.6 | 1.71 | 1.51 | 1.52 | 0.68 |
| 0.8 | 1.13 | 1.08 | 0.99 | 0.39 |
| 1.0 | 0.84 | 0.76 | 0.71 | 0.21 |

Table 2: **Average frequency error for standard count-min (CM), double-hashing with days 0-4 training and day 5 validation (DH 1), double-hashing with days 74-78 training and day 79 validation (DH 2), and the ideal learned count-min (CM-Ideal) for the 80th day prediction.**

## 6 CONCLUSIONS

In the present work we introduced a novel double-hashing algorithm for frequency estimation in data streams. The algorithm benefits from both the efficiency of the hashing-based data sketches and the accuracy provided by the learning-augmented algorithms.

Because the double-hashing algorithm is not dependent on specific features of the stream elements for identifying heavy hitters, it is applicable to a large variety of streams.

We demonstrated better accuracy of the double-hashing algorithms in estimating frequencies compared to the classical count-min sketch approach on synthetic data and on real data streams from the AOL internet queries dataset.

We proposed a continuous optimization approach that is enabled by the efficiency of the double-hashing algorithm, which does not employ additional machine learning models to identify heavy hitters. We demonstrated on the AOL data that the continuously optimized double-hashing algorithm achieves further reduction of errors in frequency estimates and provides competitive performance compared to a machine-learning augmented algorithm.

## ACKNOWLEDGMENTS

Authors would like to thank Capital One Financial Corp. for supporting this work and for allowing it to be published. We also would like to thank prof. Pyotr Indyk and Dr. Chen-Yu Hsu for providing us the AOL dataset.